\documentclass[aip,apl,reprint]{revtex4-1}

\usepackage[T1]{fontenc}
\usepackage{graphicx}
\usepackage{units}
\usepackage{amsmath,amssymb}
\usepackage{color}
\usepackage{upgreek}
\begin{document}

\title{Tuning the supercurrent distribution in parallel ballistic graphene Josephson junctions}

\author{Philipp Schmidt}
\email[]{philipp.schmidt3@rwth-aachen.de}
\affiliation{JARA-FIT and 2nd Institute of Physics, RWTH Aachen University, 52074 Aachen, Germany}
\affiliation{Peter Gr\"unberg Institute (PGI-9), Forschungszentrum J\"ulich, 52425 J\"ulich, Germany}
\author{Luca Banszerus}
\affiliation{JARA-FIT and 2nd Institute of Physics, RWTH Aachen University, 52074 Aachen, Germany}
\affiliation{Peter Gr\"unberg Institute (PGI-9), Forschungszentrum J\"ulich, 52425 J\"ulich, Germany}
\author{Benedikt Frohn}
\affiliation{JARA-FIT and 2nd Institute of Physics, RWTH Aachen University, 52074 Aachen, Germany}
\author{Stefan Blien}
\affiliation{Institute for Experimental and Applied Physics, University of Regensburg, 93040 Regensburg, Germany, EU}
\author{Kenji Watanabe}
\affiliation{Research  Center  for  Functional  Materials,  National  Institute for  Materials  Science,  1-1  Namiki,  Tsukuba  305-0044,  Japan}
\author{Takashi Taniguchi}
\affiliation{International  Center  for  Materials  Nanoarchitectonics,  National Institute  for  Materials  Science,  1-1  Namiki,  Tsukuba  305-0044,  Japan}
\author{Andreas K. Hüttel}
\affiliation{Institute for Experimental and Applied Physics, University of Regensburg, 93040 Regensburg, Germany, EU}
\author{Bernd Beschoten}
\affiliation{JARA-FIT and 2nd Institute of Physics, RWTH Aachen University, 52074 Aachen, Germany}
\author{Fabian Hassler}
\affiliation{JARA-Institute for Quantum Information, RWTH Aachen University, 52056 Aachen, Germany}
\author{Christoph Stampfer}
\affiliation{JARA-FIT and 2nd Institute of Physics, RWTH Aachen University, 52074 Aachen, Germany}
\affiliation{Peter Gr\"unberg Institute (PGI-9), Forschungszentrum J\"ulich, 52425 J\"ulich, Germany}

\date{\today}

\begin{abstract}
We report on a ballistic and fully tunable Josephson junction system consisting of two parallel ribbons of graphene in contact with superconducting MoRe.
By electrostatic gating of the two individual graphene ribbons we gain control over the real space distribution of the superconducting current density, which can be  continuously tuned between both ribbons.
We extract the respective gate dependent spatial distributions of the real space current density by employing Fourier- and Hilbert transformations of the magnetic field induced modulation of the critical current. This approach is fast and does not rely on a symmetric current profile. 
It is therefore a universally applicable tool, potentially useful for carefully adjusting Josephson junctions.

\end{abstract}


\maketitle

Josephson junctions\cite{Josephson1962}, which consist of two superconductors connected by a normal conducting material or an insulator, have been investigated for a long time, since they can be used for infrared detectors\cite{Walsh2017}, ultrafast logic circuits\cite{Polonsky1993} or sensitive magnetic flux and voltage measurements\cite{GRUNDMANN200517}.
Additionally, Josephson junctions are a powerful tool for exploring the properties of superconductors by connecting the supercurrent to the phase of the macroscopic wave function\cite{gross2016applied}.
In the last decade, Josephson junctions have also been employed as building blocks for superconducting quantum computing\cite{Huang2020, Kjaergaard2020Mar, Blais2021May}.
Substituting the normal conductor or insulator with graphene as the weak link, leads to highly tunable Josephson junctions with transparent interfaces due to the absence of a Schottky barrier\cite{Calado2015}.
In the past, graphene-based Josephson junctions\cite{Ke2016,Borzenets2016,Heersche2007,Du2008,OjedaAristizabal2009,Borzenets2011,Komatsu2012,Mizuno2013,Choi2013,Li_2018} have been investigated by tunneling spectroscopy\cite{Bretheau2017,Wang2018Sep} and have been used to study crossed Andreev reflection\cite{Park2019} and superconductivity in the quantum Hall regime\cite{Rickhaus2012Apr,Draelos2018Jun, Zhao2020Aug,Gul2022Jun}.
Besides the magnitude of the supercurrent carried over the Josephson junction, the spatial current distribution has been analyzed in the case of edge currents in graphene\cite{Zhu2017Feb,Allen2016Feb, Allen2017} and for studying topological Josephson junctions\cite{Ying2020}. Furthermore, the control and determination of the current density is important for the operation of protected Josephson rhombus chains~\cite{Bell2014Apr} or protected superconducting qubits based on tunable Josephson interferometer arrays\cite{Schrade2022}.
Here, the pairwise balance of the Josephson junctions protects the qubit against detuning or noise.
However, the real space current density is not directly accessible in electrical transport measurements.

To obtain the real space current density of two coupled Josephson junctions, an out-of-plane magnetic field has to be applied to the junction, which leads to a modulation of the critical current. This can be expressed as the magnitude of a Fourier transform of the real space current density, which, thus, can be reconstructed from the modulation of the critical current by an inverse Fourier transform\cite{Dynes1971May}.
Even though this method has been recently applied to a gated epitaxial Al-InAs Josephson junction\cite{Elfeky2021}, highly non-symmetric cases have not been studied so far.
In this work, we study a fully tunable graphene double Josephson junction formed by two parallel ribbons with superconducting molybdenum-rhenium (MoRe) contacts. By using both top and back gates, we can independently tune the supercurrent distribution between the two ribbons. 
We present results based on a reconstruction of the supercurrent from the magnetic field dependent critical current using a combination of Fourier and Hilbert transformations, which allows to extract the asymmetric supercurrent distribution.

\begin{figure*}
\includegraphics[width=0.93\linewidth]{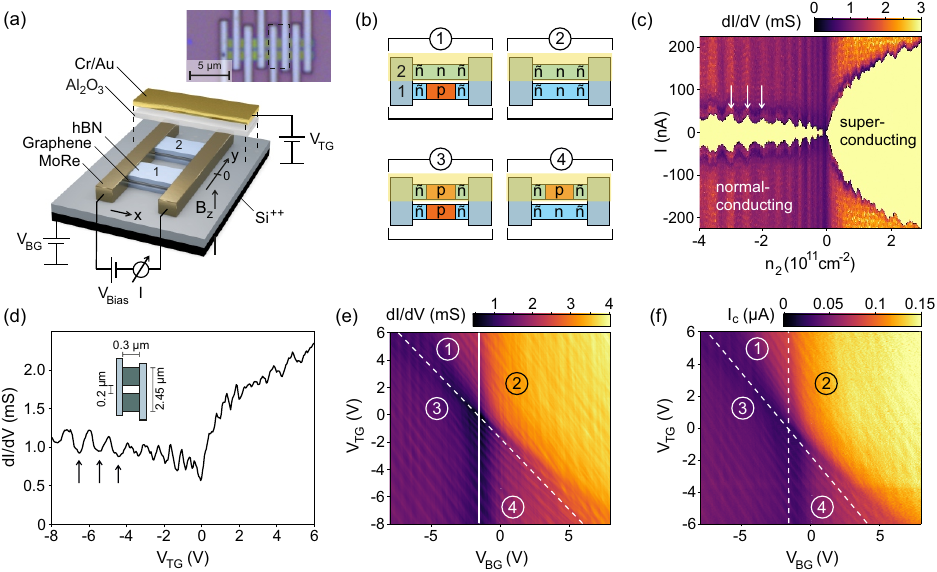}
\caption{(a) Schematics of the device structure and optical micrograph of the device. Two etched graphene ribbons (1 and 2) encapsulated in hBN are contacted with superconducting MoRe. In addition to the global Si$^{++}$ back gate, a Cr/Au top gate is deposited onto one of the ribbons (not shown in the micrograph), separated by a layer of aluminum oxide (Al$_2$O$_3$). A constant bias voltage $V_{\mathrm{Bias}}$ is applied and the resulting current $I$ is measured.
(b) Schematic of the doping configurations accessible with the two gates. The MoRe contacts induce a persistent $\mathrm{\Tilde{n}}$-doping at their graphene interfaces.
(c) Differential conductance as function of charge carrier density $n_2$ in the top-gated graphene ribbon and current ($T \approx 10\,\mathrm{mK}$). 
The superconducting regime around $I=0$ is well visible. For p-doping, i.e. $n<0$, Fabry-Pérot oscillations (see vertical arrows) are visible in the critical current.
(d) Normal state conductance as function of top gate voltage $V_{\mathrm{TG}}$ with reference to the charge neutrality point. 
The Fabry-Pérot oscillations are also marked by arrows. 
The inset shows a sketch of the device with the relevant dimensions.
(e) The normal state conductance vs applied top- and back gate voltages showing four electrostatic doping regimes of the graphene ribbons. 
The white vertical line correspond to the position of the line-trace shown in panel (d).
(f) The gate dependent critical currents show a tuning similar to the normal state conductance in panel (e).}
\label{fig:fig1}
\end{figure*}

A schematic and an optical micrograph of our device is shown in Fig.~1a. It consists of graphene grown by chemical vapor deposition (CVD), which is encapsulated in hexagonal boron nitride (hBN) crystals by dry transfer van der Waals stacking.\cite{Wang2013, Banszerus2015Jul, Banszerus2016Feb} The stack is etched into two parallel ribbons by CF$_4$/O$_2$ reactive ion etching (RIE) through a polymethyl methacrylate (PMMA) resist mask, which has been patterned by standard electron beam lithography to a width of $w=1.1\,\mathrm{\upmu m}$ with a separation of $\delta=0.2\,\mathrm{\upmu m}$ between the two ribbons. The graphene is electrically contacted to superconducting MoRe electrodes, fabricated by sputter deposition defining the junction length of $L=0.3\,\mathrm{\upmu m}$.
Separated by 300~nm SiO$_2$ gate dielectric, the device is placed on a highly doped silicon substrate acting as a back gate. Additionally, a gold top gate is deposited onto one of the graphene ribbons, while the whole device is protected by an insulating layer of Al$_2$O$_3$ grown by atomic layer deposition (ALD). This allows us to tune the charge carrier density of the two graphene ribbons independently since one of them is only affected by the back gate (BG) leading to a charge carrier density of $n_1=\alpha_{\mathrm{BG}}(V_{\mathrm{BG}}-V_{\mathrm{BG}}^0)$ while the other ribbon is tuned by the combination of both top gate (TG) and back gate voltages $n_2=\alpha_{\mathrm{BG}}(V_{\mathrm{BG}}-V_{\mathrm{BG}}^0)+\alpha_{\mathrm{TG}}(V_{\mathrm{TG}}-V_{\mathrm{TG}}^0)$, where $\alpha_{\mathrm{BG}}$ and $\alpha_{\mathrm{TG}}$ are the respective gate lever arms, which are proportional to the capacitive coupling of the respective gate to the graphene. 
The gate settings $V_{\mathrm{BG}}^0=-1.3\,\mathrm{V}$ and $V_{\mathrm{TG}}^0=-0.25\,\mathrm{V}$ account for a constant shift of the charge neutrality point and the gate lever arms are estimated to be  $\alpha_{\mathrm{BG}}=7.0\times 10^{10} \, \mathrm{V^{-1}cm^{-2}}$ and $\alpha_{\mathrm{TG}}=4.9\times 10^{10} \, \mathrm{V^{-1}cm^{-2}}$.
With this individual gate tuning, the device can be tuned into four regimes defined by the polarities of the two ribbons, see Fig.~1b: p|n (1), n|n (2), p|p (3) and n|p (4).
All measurements were performed in a He$^3$/He$^4$ dilution refrigerator with a base temperature of around 10\,mK. A constant parasitic resistance of $R_0=2.55\,\mathrm{k\Omega}$ arises from the wiring and the filters.

In Fig.~1c we show the differential conductance $dI/dV$ of the device as function of the current $I$ and charge carrier density $n_2$, which is tuned by the top gate voltage while the back gate voltage is set to the charge neutrality point ($V_{\mathrm{BG}}=-1.3\,\mathrm{V}$). A superconducting regime (yellow color) is present symmetric around $I=0$ for both electron ($n>0$) and hole ($n<0$) doping and can be distinguished from the normal conducting regime at larger current values.
The current at which the transition between these two regimes occurs defines the critical current $I_\mathrm{c}$. Here it is important to note that no influence of the current sweep direction on the critical current has been observed.

Oscillations of the differential conductance and the critical current that are observed in the p-doped region ($n<0$) can be attributed to Fabry-Pérot (FP) interference\cite{Rickhaus2013Aug, BenShalom2016}.
They originate from the $\mathrm{\Tilde{n}}$-p-$\mathrm{\Tilde{n}}$ graphene cavity in the double gated ribbon, which is formed by the MoRe yielding a highly local $\mathrm{\Tilde{n}}$ doping in the graphene nearby the contacts (see Fig.~1b).
The critical current becomes largest for n-doping ($n>0$) since there is no p-$\mathrm{\Tilde{n}}$ junction, which increases the resistance. Since the carrier density is only weakly tuned in the outer graphene parts next to the MoRe contacts, there is no cavity formed, as seen by the suppression of FP oscillations in this regime. 
The FP oscillations are further explored in Fig.~1d, where we plot the normal state conductance, measured at a bias voltage of $1\,\mathrm{mV}$ to keep the junction in the normal state regime, at fixed back gate voltage and varying top gate voltage. Analyzing the FP oscillations (see Supplementary Material for details) for both ribbons results in an extracted cavity length of $L_1=314\,\mathrm{nm}\pm 50\,\mathrm{nm}$ and $L_2=335\,\mathrm{nm}\pm 15\,\mathrm{nm}$ which fits well to the lithographic length of the channel (see inset in Fig.~1d for sample dimensions), in agreement with ballistic and phase coherent transport.
In Fig.~1e, we show the normal state conductance of the device as function of top- and back gate voltages. The four possible doping configurations sketched in Fig.~1b can be identified (see labels in Figs.~1e,f).
The top gate voltage only changes the charge carrier density of the respective graphene ribbon underneath, while the back gate voltage influences both ribbons.
Configuration (1) refers to the situation of a $\mathrm{\Tilde{n}}$-p-$\mathrm{\Tilde{n}}$ junction in one ribbon, while the other (underneath the top gate) is completely n-doped. The complementary configuration, denoted as (4), has the $\mathrm{\Tilde{n}}$-p-$\mathrm{\Tilde{n}}$ junction underneath the top gate. Configuration (3) refers to the case of both ribbons tuned to a $\mathrm{\Tilde{n}}$-p-$\mathrm{\Tilde{n}}$ junction. 
Here, a combination of the FP oscillations by the two gates can be seen.
Finally, configuration (2) describes the case of unipolar n-type doping for both ribbons and thus the absence of any p-$\mathrm{\Tilde{n}}$ junctions, which results in an enhanced conductance.
Reducing the bias voltage and thus entering the superconducting state enables us to extract the critical current $I_\mathrm{c}$. The critical current is plotted as function of top- and back gate voltages in Fig.~1f. The map resembles most features of the conductance map in Fig.~1e.
Combining the critical current and the normal state resistance gives an $I_\mathrm{c} R_\mathrm{N}$ product between $30\,\mathrm{\upmu V}$ and $60\,\mathrm{\upmu V}$, comparable to results in the literature\cite{Li2018Feb}.

\begin{figure}
\includegraphics[width=0.98\linewidth]{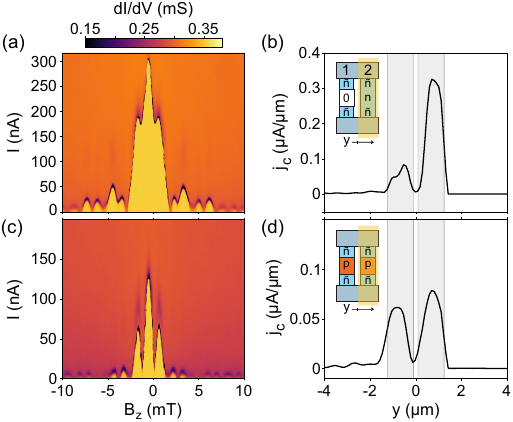}
\caption{(a) Differential conductance as function of current and out-of-plane magnetic field for the gating configuration of one conductive, n-doped graphene ribbon while the other is tuned to charge neutrality (see inset of panel (b)). The critical current shows a Fraunhofer-like modulation pattern. (b) The extracted real space current density confirms the expected asymmetric current distribution by the pronounced peak at $y>0$. (c) Differential conductance map for the case where both ribbons are p-doped. The magnetic field induced modulation of the critical current changes to a more SQUID-like behavior. (d) The resulting real space current density shows two peaks with almost equal intensity, proving the symmetric current distribution in the two ribbons.}
\label{fig:fig2}
\end{figure}

Next we focus on the magnetic field dependence of the critical current, which will allow to reconstruct its spatial distribution.
We measure the differential conductance while varying both the current $I$ and the out-of-plane magnetic field $B_\mathrm{z}$. The magnetic field induces a phase difference between the two bulk superconductors, leading to a modulation of the critical current $I_\mathrm{c}$ \cite{Rowell1963Sep} as shown in Fig.~2a for the case of one graphene ribbon being conductive ($\mathrm{\Tilde{n}}$-n-$\mathrm{\Tilde{n}}$) while the other is tuned near the charge neutrality point ($\mathrm{\Tilde{n}}$-0-$\mathrm{\Tilde{n}}$) (see inset of Fig.~2b).
To gain information on the real space superconducting current density distribution, $j_c$, we perform an inverse Fourier transformation (FT) of the extracted $I_\mathrm{c}(B)$ combined with a Hilbert transformation to reconstruct the phase of the signal as introduced by Dynes and Fulton \cite{Dynes1971May}, see Supplementary Material for details.
Note that the reconstructed data leaves the freedom of an arbitrary offset in the position $y$ together with the sign of $y$ (which corresponds to mirroring around $y=0$). We correct the reconstructed data such that position $y=0$ correspond to the center of the two graphene ribbons and flip the $y$-axis such that the major peak corresponds to the  ribbon with the larger conductance. Here, the junction tuned by the back gate only, is located at negative position ($y<0$) whereas the top gated junction is located at positive position ($y>0$).
The extracted current density distribution in Fig.~2b of the measurement shown in Fig.~2a visualizes the asymmetric supercurrent distribution in this configuration with the largest critical currents along the ($\mathrm{\Tilde{n}}$-n-$\mathrm{\Tilde{n}}$) ribbon at $y>0$.
Changing the gate voltages to the symmetric configuration (see configuration (3) in Fig.~1b), where both graphene ribbons are p-doped shows a different modulation pattern (see Fig.~2c) and the extracted supercurrent density shows indeed an even current distribution between the two ribbons (see Fig.~2d).
We use the width of these current peaks to adapt the $y$-axis of the reconstructed data to $y=\Phi_0/(Bt_\mathrm{B})$ with the magnetic flux quantum $\Phi_0=h/(2e)$ and $t_{\mathrm{B}}=L+\lambda_1+\lambda_2$. Taking into account the length obtained by the analysis of FP oscillations of $L=335\,\mathrm{nm}$ results in a London penetration depth of $\lambda_{1/2}\approx 330\,\mathrm{nm}$.

\begin{figure}
\includegraphics[width=0.98\linewidth]{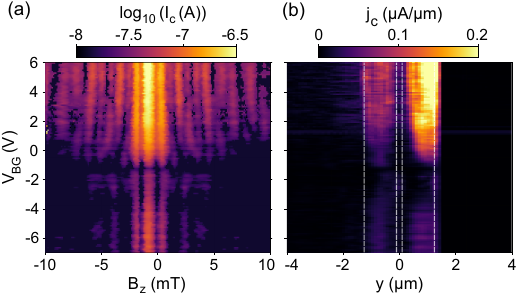}
\caption{(a) Modulation of the critical current --- in logarithmic scale --- as function of out-of-plane magnetic field and back gate voltage at fixed $V_\mathrm{TG}=6.3\,\mathrm{V}$. The modulation pattern changes with applied gate voltage. (b) Line-wise reconstruction from the data in panel (a) resulting in the real space position and gate voltage dependent superconducting current density distribution. The current density shows non-symmetric distribution as well as a symmetrical influence of the back gate on the two graphene ribbons.}
\label{fig:fig3}
\end{figure}

To measure the critical current directly instead of extracting it from multidimensional measurements, we used a home-built circuit which detects the peak in differential resistance when sweeping the applied current (see Supplementary Material for details). 
This allows an enhanced measurement speed by a factor of 60 and opens the door to measure the modulation patterns as a function of the applied gate voltages. 
In Fig.~3a, we show the logarithm of the critical current as function of magnetic field and $V_\mathrm{{BG}}$. 
Again, the magnetic field induced modulation that depends on the applied back gate voltage is visible by the features of brighter colors. 
Even the modulation features of low intensity at higher magnetic fields can be seen.
Strikingly, the critical current is lower for negative back gate voltages, because of the resulting $\mathrm{\Tilde{n}}$-p-$\mathrm{\Tilde{n}}$-junction, which increases the normal resistance which is less pronounced for positive voltages.
We perform the reconstruction for each line and show the gate dependent real space current distribution in Fig.~3b. 
The most important features are the two distinctive areas with high current densities at $y\approx \pm 1\,\mathrm{\upmu m}$ which correspond to the two graphene ribbons. 
It can be seen that the back gate indeed tunes both junctions and thus changes the current densities for the areas at $y<0$ and $y>0$. 
The asymmetry is caused by the top gate influencing only ribbon~2 (see Figs.~1a,b).

\begin{figure*}
\includegraphics[width=0.99\linewidth]{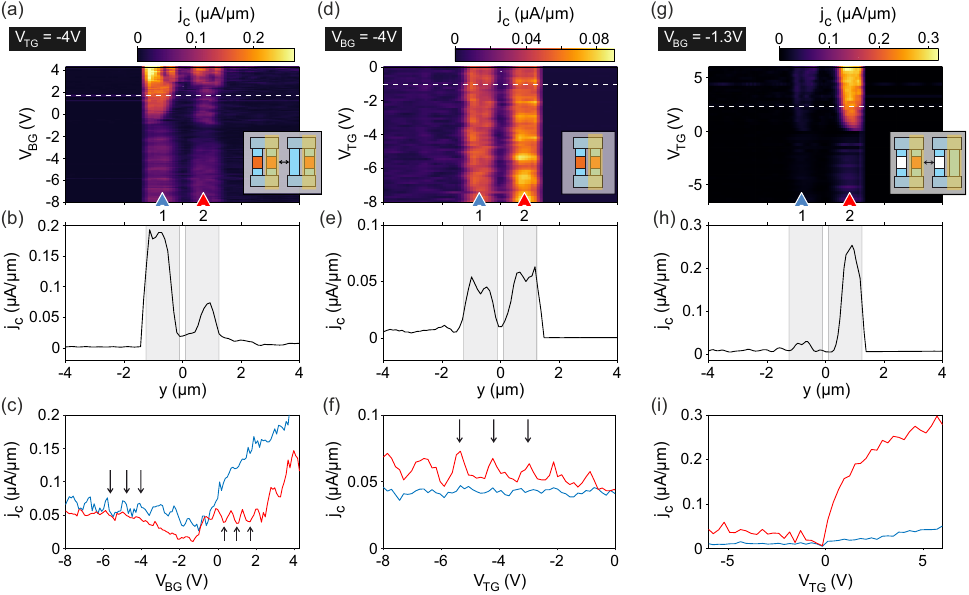}
\caption{
(a) Superconducting current density as function of real space position and back gate voltage for a doping configuration where both ribbons are tuned by the back gate with fixed top gate voltage. The inset illustrates the respective doping configuration changing from p|p to n|p with increasing gate voltage.
(b) Line profile of the superconducting current density along the white dashed line in panel (a). The asymmetric current distribution between the two ribbons can be well identified. The grey areas correspond to the lithographic defined positions of the graphene ribbons 1 and 2.
(c) Line profiles of the current densities along the gate axis at the left and right junction indicated by blue and red triangles in panel (a), respectively. The Fabry-Pérot oscillations, marked by the black arrows, can also be seen in the current density.
(d-f) Similar as (a-c) but the top gate voltage is varied while both ribbons are p-doped leading to a symmetric current distribution.
(g-i) Similar as (d-f) but ribbon 1 is tuned to the charge neutrality point. Therefore, almost all the current flows through ribbon 2 leading to a highly asymmetric current distribution.
}
\label{fig:fig4}
\end{figure*}

We now focus on the real space superconducting current densities for different electrostatic configurations in more detail.
In the first configuration, presented in Fig.~4a, most current flows through ribbon 1 that is not covered by the top gate at $y<0$. 
In Fig.~4a the superconducting current density is plotted as function of real space position and back gate voltage. 
The back gate voltage tunes the bulk of ribbon 1 from n-doping at $V_\mathrm{BG}>0$ to p-doping, resulting in a $\mathrm{\Tilde{n}}$-p-$\mathrm{\Tilde{n}}$ configuration at $V_\mathrm{BG}<0$ while the fixed top gate 
voltage keeps the bulk of ribbon~2 for all $V_\mathrm{BG}$ values p-doped (i.e. in the $\mathrm{\Tilde{n}}$-p-$\mathrm{\Tilde{n}}$ configuration), as shown by the inset in Fig.~4a. 
The map shows two vertical features of high current density that can be assigned to the supercurrents running through the two ribbons. 
The highest current density is present in case of n-doping ($V_\mathrm{BG}>0$) for $y<0$, resulting from the higher normal conductance of the $\mathrm{\Tilde{n}}$-n-$\mathrm{\Tilde{n}}$ ribbon. 
In Fig.~4b we show a line-cut along the dashed line in Fig.~4a where the asymmetry is well visible as the peak at $y<0$ is more pronounced.
Additionally, line-traces of the current density as function of $V_\mathrm{BG}$ extracted at the center positions of the two ribbons (see blue and red triangles in Fig.~4a) are shown in Fig.~4c. 
While ribbon 1 (blue trace) shows a clear transition from reduced current density with FP oscillations at $V_\mathrm{BG}<0$ to an increasing current density for $V_\mathrm{BG}>0$, ribbon 2 (red trace) shows FP oscillations even for positive $V_\mathrm{BG}$ values (see black arrows). This gives us an unambiguous way to identify the peaks in the current density as the junction with or without top gate.

A similar analysis is performed for the configuration where we fix $V_\mathrm{BG}$ and tune $V_\mathrm{TG}$ so that both graphene ribbons are p-doped (see Fig. 4d-f). 
Here, the gate voltage and position dependent current density (Fig.~4d) shows a rather symmetric distribution between the two ribbons visualized by the two parallel vertical strips of high current density for all measured top gate voltages. The even distribution can also be seen in the line-cut at $V_{\mathrm{TG}}\approx -1\,\mathrm{V}$, see Fig.~4e. 
Moreover, we observe an oscillatory modulation of the real space current density with respect to the gate voltage due to FP interference in the $\mathrm{\Tilde{n}}$-p-$\mathrm{\Tilde{n}}$ cavity. This effect becomes  more visible in the gate dependent line-cuts at the position of the two ribbons as indicated by the vertical arrows (see Fig.~4f).
In this configuration the FP oscillation is only present for the ribbon underneath the top gate (ribbon 2), since $V_\mathrm{BG}$ is kept constant and thus the tuning of charge carrier density leading to the FP oscillation is asymmetric.

Finally, in Figs.~4g-i we show a third configuration where only the graphene ribbon underneath the top gate is conductive, while the other is tuned to its charge neutrality point by the back gate. 
Consequently, the extracted current density shows a highly asymmetric distribution among the ribbons with almost no supercurrents at $y<0$, see Fig.~4g. 
This also becomes clear in the line-cut at $V_\mathrm{{TG}}\approx 2.4\,\mathrm{V}$ (Fig.~4h), where only one significant peak at $y>0$ can be identified. Furthermore, the current suppression in the case of a $\mathrm{\Tilde{n}}$-p-$\mathrm{\Tilde{n}}$ junction can be seen in Fig.~4i when comparing the current density at negative and positive gate voltage. This fits well to the observed difference in critical current for n- and p-doping (compare to Fig.~1f).

In summary, we presented a fully tunable graphene-based double Josephson junction consisting of two parallel graphene ribbons contacted by MoRe. We showed a remarkable control over these junctions and are able to individually tune the charge carrier densities of both ribbons. 
This allows us to continuously control the distribution of the superconducting current density in the two ribbons independently.
Through ballistic transport, ensured by the dry transfer fabrication method of the device, we were able to observe FP oscillations. From these oscillations, we could determine the length of the electrostatic $\mathrm{\Tilde{n}}$-p-$\mathrm{\Tilde{n}}$ cavity between the superconducting contacts.
In order to test the tuning capabilities of the graphene Josephson junction we measured the critical current depending on magnetic field and gate voltages leading to the magnetic field induced modulation of the critical current. 
Most interestingly, we could map the current density distributions in real space of the individual graphene ribbons, highlighting the FP oscillations present in the individual ribbons and proving the excellent gate tunability of the junctions from an even to a fully asymmetric current distribution.
This opens the door of controlling and monitoring the current densities in complex Josephson interferometer circuits potentially leading to protected superconducting qubits\cite{Schrade2022}.

\begin{acknowledgments}
We thank U. Wichmann for help with the measurement electronics.
This project has received funding from the Deutsche Forschungsgemeinschaft (DFG, German Research Foundation) under Germany’s Excellence Strategy – Cluster of Excellence Matter and Light for Quantum Computing (ML4Q) EXC 2004/1 – 390534769, the European Union’s Horizon 2020 research and innovation programme under grant agreement No. 881603 (Graphene Flagship) and from the European  Research Council (ERC) (grant agreement No.820254)), and the Helmholtz Nano Facility \cite{Albrecht2017May}. K.W. and T.T. acknowledge support from JSPS KAKENHI (Grant Numbers 19H05790, 20H00354 and 21H05233) and A.K.H. acknowledges support from the DFG (Hu 1808/4-1, project id 438638106).

\end{acknowledgments}

\textbf{Data availability} The data supporting the findings are available in a Zenodo repository under accession
code https://doi.org/10.5281/zenodo.7632431.

\end{document}


\title{\bf Supplemental Material for: \\ Tuning the supercurrent distribution in parallel ballistic graphene Josephson junctions}

\author{Philipp Schmidt}
\email[]{philipp.schmidt3@rwth-aachen.de}
\affiliation{JARA-FIT and 2nd Institute of Physics, RWTH Aachen University, 52074 Aachen, Germany}
\affiliation{Peter Gr\"unberg Institute (PGI-9), Forschungszentrum J\"ulich, 52425 J\"ulich, Germany}
\author{Luca Banszerus}
\affiliation{JARA-FIT and 2nd Institute of Physics, RWTH Aachen University, 52074 Aachen, Germany}
\affiliation{Peter Gr\"unberg Institute (PGI-9), Forschungszentrum J\"ulich, 52425 J\"ulich, Germany}
\author{Benedikt Frohn}
\affiliation{JARA-FIT and 2nd Institute of Physics, RWTH Aachen University, 52074 Aachen, Germany}
\author{Stefan Blien}
\affiliation{Institute for Experimental and Applied Physics, University of Regensburg, 93040 Regensburg, Germany, EU}
\author{Kenji Watanabe}
\affiliation{Research  Center  for  Functional  Materials,  National  Institute for  Materials  Science,  1-1  Namiki,  Tsukuba  305-0044,  Japan}
\author{Takashi Taniguchi}
\affiliation{International  Center  for  Materials  Nanoarchitectonics,  National Institute  for  Materials  Science,  1-1  Namiki,  Tsukuba  305-0044,  Japan}
\author{Andreas K. Hüttel}
\affiliation{Institute for Experimental and Applied Physics, University of Regensburg, 93040 Regensburg, Germany, EU}
\author{Bernd Beschoten}
\affiliation{JARA-FIT and 2nd Institute of Physics, RWTH Aachen University, 52074 Aachen, Germany}
\author{Fabian Hassler}
\affiliation{JARA-Institute for Quantum Information, RWTH Aachen University, 52056 Aachen, Germany}
\author{Christoph Stampfer}
\affiliation{JARA-FIT and 2nd Institute of Physics, RWTH Aachen University, 52074 Aachen, Germany}
\affiliation{Peter Gr\"unberg Institute (PGI-9), Forschungszentrum J\"ulich, 52425 J\"ulich, Germany}

\maketitle

\section{Extraction of the supercurrent density distribution}
The formalism to extract the current density distribution from the magnetic field induced modulation of the critical current closely follows the method introduced by Dynes and Fulton~\cite{Dynes1971May} which is briefly discussed here.

In general, the magnetic field dependent critical current $I_c$ of a short Josephson junction can be expressed as
\begin{equation}\label{eq:ic_dynes}
    I_c(B)=\left| \int_{-\infty}^\infty dy\, j_c(y) e^{2\pi \mathrm{i} (2\lambda +L)By/\Phi_0} \right|,
\end{equation}
where $y$ denotes the coordinate along the width of the superconducting contacts (see Fig.~1 of main text), $\Phi_0=h/(2e)$ is the magnetic flux quantum, $\lambda$ is the magnetic penetration depth and $L$ is the distance between the contacts, which is determined by analyzing Fabry-Pérot oscillations. 
Thus, $I_c$ is given by the complex Fourier transform of the real space current density $j_c(y)$.
To compute the current density from the measured critical current modulation, we employ an inverse Fourier transform:
\begin{equation}
    j_c(y) = \frac{1}{2\pi} \int_{-\infty}^\infty d\beta \, I_c(\beta)e^{\mathrm{i}\theta}\, e^{-\mathrm{i}\beta y},
\end{equation}
using a normalized magnetic field $\beta=2\pi (2\lambda +L)B/\Phi_0$.
To account for the general case, where no symmetry of the current density is provided, the phase $\theta$ is introduced, which is not directly measured since $I_c$ is just the magnitude of the complex Fourier transform.
The determination of $\theta$ from the knowledge of $I_c$ is the problem of phase-reconstruction, well-studied in signal processing. Assuming the minimal-phase-shift property \cite{Papoulis},
the phase $\theta$ is linked to the critical current $I_c$ by a Hilbert transform
\begin{equation}
    \theta(\beta) = \frac{1}{2\pi} \int_{-\infty}^\infty d\Tilde{\beta} \,\frac{\ln{I_c(\Tilde{\beta})}-\ln{I_c(\beta)} }{\beta^2 -\Tilde{\beta}^2}.
\end{equation}
From $I_c$ and $\theta$ the real space current density can then be reconstructed. Instead of numerically computing this integral we use a discrete Hilbert transform.
The Python implementation relating the critical current \texttt{ic} (as a function of the magnetic field) to the current distribution \texttt{dist} is given by the short program
\begin{lstlisting}[language=Python]
import numpy as np
from scipy import signal, fft
dist = fft.fftshift(fft.ifft(np.exp(signal.hilbert(np.log(ic)))))
\end{lstlisting}

\section{Measuring the critical current}

\begin{figure}
	\centering
	\includegraphics[width=0.95\linewidth]{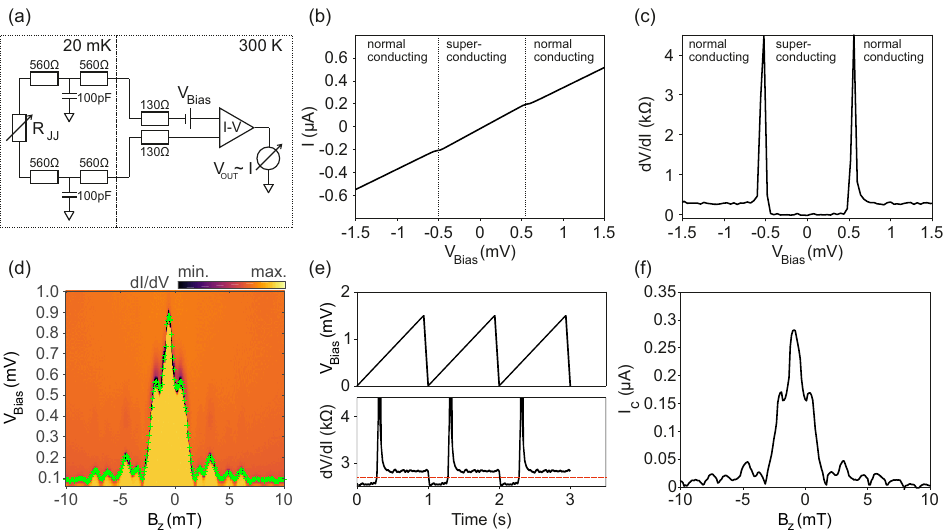}
	\caption{(a) Schematic of the two-terminal measurement configuration showing the sample resistance $R_{\mathrm{JJ}}$ that is tuned by the gates, the cryogenic RC filters and the room temperature electronics. The filters and cables lead to a total series resistance of $2.55\,\mathrm{k\Omega}$. Differential resistance measurements are performed by standard lock-in techniques. (b) A typical $I-V$ curve showing the superconducting regime by a change of the slope. (c) Differential resistance after subtracting the $2.55\,\mathrm{k\Omega}$ series resistance. (d) Differential conductance as function of applied bias voltage and magnetic field. The numerically extracted edge of the superconducting regime is marked by green crosses. (e) Applied bias voltage and differential resistance at the lock-in amplifier as function of time. A self-made electronic circuit is used to detect the a strong resistance change at the trigger level indicated as dashed red line by relating it to the applied bias voltage. (f) Extracted critical current as function of magnetic field resulting from the fast sensing method introduced in (e).}
	\label{fig:methods}
\end{figure}   

\subsection{Experimental setup}
Our measurement scheme to measure the sample resistance and thus determine the critical current is based on a two-terminal configuration where we apply a bias voltage and measure the corresponding current by an $I/V$ converter, see Fig.~\ref{fig:methods}a. In addition to the sample resistance, which is tuned by the gate voltages, the wiring and filtering (i.e. cryogenic RC filters and copper powder filters) of our setup causes a parasitic series resistance of $2.55\,\mathrm{k\Omega}$.
A typical $I-V$ curve is shown in Fig.~\ref{fig:methods}b. A transition between the superconducting and the normal conducting state can be identified at bias voltages of $V_{Bias}\approx \pm 0.5\,\mathrm{mV}$ as a change of the slope even if the series resistance dominates the overall linear behavior.
Additionally, we employ a standard low frequency lock-in technique to measure the differential resistance of the device. The differential resistance as function of applied bias voltage with $R_0$ subtracted shows two distinct levels corresponding to the two states and resistance peaks at the transition, see Fig.~\ref{fig:methods}c.

\subsection{Determination of gate lever arms}
The back gate lever arm $\alpha_{\mathrm{BG}}$ is determined by the geometry of the device to $\alpha_{\mathrm{BG}}=\epsilon_0 \epsilon_r / ( ed )=7.0\times 10^{10} \, \mathrm{V^{-1}cm^{-2}}$ with $\epsilon_r=3.8$ and gate distance $d=300\,\mathrm{nm}$ while the top gate lever arm is determined relative to $\alpha_{\mathrm{BG}}$ by using the slope of the diagonal line in Fig.~1e to $\alpha_{\mathrm{TG}} = 0.7 \times \alpha_{\mathrm{BG}}$.

\subsection{Extraction of the critical current}
To measure the critical current, we link the bias voltage $V_{Bias,c}$ at a strong resistance/conductance change to the critical current either by $I_c = V_{Bias,c}/R_0$,
or by the direct measurement of the current at that bias voltage $I_c = I(V_{Bias,c})$.
This can be done numerically as shown in Fig.~\ref{fig:methods}d where we plot the measured differential conductance as function of applied bias voltage and magnetic field. For each magnetic field value we determine $V_{Bias,c}$ (marked with green crosses).
On the other hand, for multi-dimensional measurement, i.e. capture the magnetic field induced modulation of $I_c$ as function of gate voltages, this method becomes very inefficient. Instead, we use a self-made electronic circuit that automatically ramps the bias voltage and directly measures the critical bias voltage upon a strong resistance change triggered by an adjustable threshold. The working principle of this method is shown in Fig.~\ref{fig:methods}e.
The resulting critical current shows no significant difference to the numerically extracted values, see Fig.~\ref{fig:methods}f.

\section{Analysis of Fabry-Pérot oscillations}

\begin{figure}
	\centering
	\includegraphics[width=0.95\linewidth]{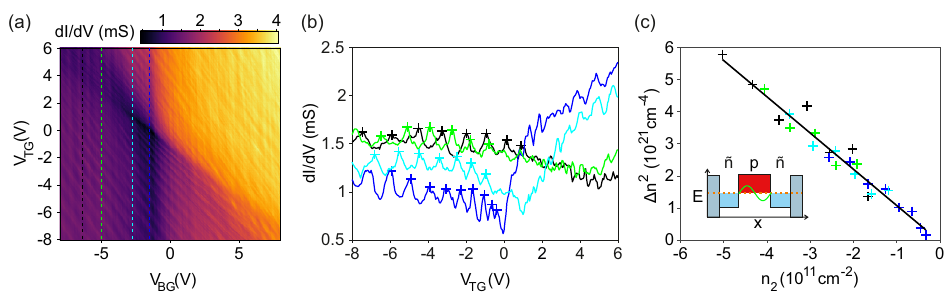}
	\caption{Analysis of Fabry-Pérot oscillations. (a) Normal state conductance as function of top- and back gate. The Fabry-Pérot oscillations are clearly visible for p-doping, i.e. negative gate voltages. (b) Conductance traces as function of top gate voltage with respect to the charge neutrality point extracted at different back gate voltages indicated by dashed lines in panel (a). (c) Squared distance of consecutive oscillation maxima marked in panel (b)  as function of charge carrier density $n_2$ at ribbon 2. Inset: Schematic of standing waves in the $\mathrm{\Tilde{n}}$-p$\mathrm{\Tilde{n}}$ cavity.}
	\label{fig:fpo}
\end{figure}   

Since the MoRe contacts induce a $\mathrm{\Tilde{n}}$-doping of the graphene in their vicinity, an electronic $\mathrm{\Tilde{n}}$-p-$\mathrm{\Tilde{n}}$ cavity is formed if the ``bulk'' graphene is p-doped by the gates. This gives rise to Fabry-Pérot oscillations of the conductance. We analyze the position of the oscillation maxima to extract the cavity length.
The interference condition for the $m$-th conductance maximum is given by
\begin{equation}
    L = m\frac{\lambda_F}{2}
\end{equation}
with cavity length $L$ and Fermi wavelength $\lambda_F$. Using the charge carrier density dependent Fermi wavelength of graphene $\lambda_F=2\sqrt{\pi /n}$ results in
\begin{equation}
    L = m \frac{\sqrt{\pi}}{\sqrt{n}}.
\end{equation}
Since in practice the maximum index $m$ is not known, we analyze pairs of consecutive maxima at position $n_0$ with distance $\Delta n$ and perform a Taylor-series expansion of the interference condition leading to
\begin{equation}
    L = \frac{2\sqrt{\pi n_0}}{\Delta n}.
\end{equation}
For this analysis, we measure the conductance as function of top- and back gate (Fig.~\ref{fig:fpo}a). For ribbon 2 we extract the conductance as function of effective top gate voltage for fixed back gate voltages, see Fig.~\ref{fig:fpo}b. The gate voltages of the oscillation maxima are extracted and converted to charge carrier densities $n_0$. Next, we plot the squared distances $\Delta n ^2$ as function of $n_0$ and fit a line to the data. Ultimately, the slope $s$ of that line gives the cavity length by
\begin{equation}
    L_2=\sqrt{\frac{4\pi}{|s|}}=335\pm 15 \,\mathrm{nm},
\end{equation}
which fits well to the lithographic length of the device.
A similar analysis of the conductance oscillations only influenced by $V_{BG}$ (vertical features in Fig.~\ref{fig:fpo}a) results in a cavity length for ribbon 1 of $L_1=314\pm 50 \,\mathrm{nm}$.

Additionally, the Fabry-Pérot oscillations visible in the extracted real space current densities (see Fig.~4 of the main text) were evaluated in the same way and result in cavity lengths of $L_1^J=392\pm 45 \,\mathrm{nm}$ and $L_2^J=382\pm 105\,\mathrm{nm}$.

\section{Influence of graphene's skewed current-phase relation}

\begin{figure}
	\centering
	\includegraphics[width=0.95\linewidth]{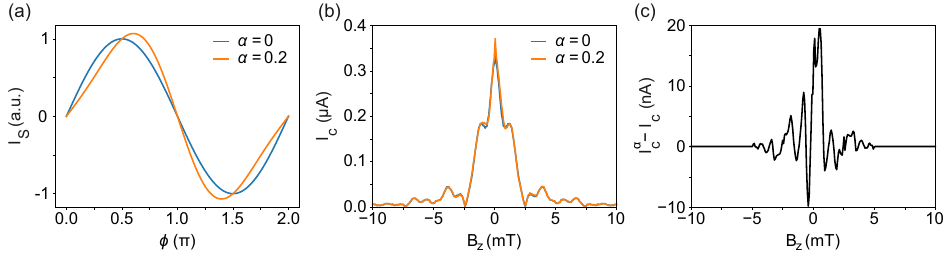}
	\caption{(a) Model of the current-phase relation for a `normal' Josephson junction ($\alpha =0$) showing a sinusoidal behavior and for a graphene Josephson junction with induced skewness ($\alpha=0.2$) (b) Computed modulation of the critical current as function of magnetic field for the two cases shown in (a). (c) Difference of the modulation pattern with and without taking  the skewed current-phase relation into account.}
	\label{fig:cpr}
\end{figure}

One of the main figures of a Josephson junction is the current-phase relation (CPR) that relates the supercurrent $I_s$ through the junction to the phase difference of the superconductors $\phi$. 
The introduced method to extract the real space current density by Fourier transform relies on a purely sinusoidal current-phase relation $I_s=I_c\sin{\phi}$ whereas deviations from this sinusoidal relation are found for graphene~\cite{Manjarres2020, Nanda2017, English2016, Lee2018}.
To model this deviation, we extend the CPR by a first order correction $I_s=I_c(\sin{\phi} - \alpha \sin{2\phi})$, with $\alpha \ll1$, leading to a skewed CPR, see Fig.~\ref{fig:cpr}a. Similar to other experimental and theoretical works, we define the skewness factor $S=(2\phi_\text{max}/\pi)-1 \approx 4\alpha/\pi$ by the phase $\phi_\text{max}\approx \tfrac12\pi + 4\alpha$ for which the supercurrent is maximized. Thus, a skewness factor of $S\approx0.25$, as reported in literature, can be achieved by a first order correction of $\alpha \approx 0.2$.
Due to the skewed current-phase relationship, the Dynes and Fulton expression \eqref{eq:ic_dynes} for the critical current is modified and we obtain
\begin{equation}\label{eq:ic_dynes2}
    I_c(B)=\max_{\varphi_0}\,\mathop{\rm Im} \int_{-\infty}^\infty dy\, j_c(y)\Bigl[e^{\mathrm{i}\beta y - i \varphi_0} -\alpha e^{2\mathrm{i}\beta y - 2i \varphi_0} \Bigr] .
\end{equation}
To lowest order in $\alpha$, the maximum is achieved at $\varphi_0 = \theta - \tfrac12\pi$.
In order to assess the influence of a potential skewed current-phase relation, we compare the critical current \eqref{eq:ic_dynes2} from the extracted current-density profile $j_c(y)$ in Fig.~\ref{fig:cpr}(b) and~\ref{fig:cpr}(c).
Except for deviations of the order of $10\,\text{nA}$, the critical currents agree. Thus, skewing of the CPR is not important for the physics we discuss in our manuscript.

\bibliographystyle{aipnum4-1}